\documentclass{PoS}

\title{Construction of a Medium-Sized Schwarzschild-Couder Telescope
  for the Cherenkov Telescope Array: Implementation of the Cherenkov-Camera Data Acquisition System}

\ShortTitle{Data Acquisition System of the MST-SCT for CTA}

\author{\speaker{M.~Santander},$^a$ J.~Buckley,$^b$ B.~Humensky,$^c$ R.~Mukherjee,$^a$ for the CTA Consortium\\
\llap{$^a$} Department of Physics and Astronomy \\
Barnard College, Columbia University, New York, NY, USA \\
\llap{$^b$} Department of Physics \\
Washington University in Saint Louis, Saint Louis, MO, USA \\
\llap{$^c$} Deparment of Physics \\
Columbia University, New York, NY, USA \\
E-mail: \email{santander@nevis.columbia.edu}, \email{buckley@physics.wustl.edu}, \email{humensky@nevis.columbia.edu},  \email{muk@astro.columbia.edu}}

\abstract{A medium-sized Schwarzchild-Couder Telescope (SCT) is being developed as a possible extension for the Cherenkov Telescope Array (CTA). The Cherenkov camera of the telescope is designed to have 11328 silicon photomultiplier pixels capable of capturing high-resolution images of air showers in the atmosphere. The combination of the large number of pixels and the high trigger rate (> 5 kHz) expected for this telescope results in a multi-Gbps data throughput. This sets challenging requirements on the design and performance of a data acquisition system for processing and storing this data.

A prototype SCT (pSCT) with a partial camera containing 1600 pixels, covering a field of view of 2.5 x 2.5 square degrees, is being assembled at the F.L. Whipple Observatory.  We present the design and current status of the SCT data acquisition system.}

\FullConference{The 34th International Cosmic Ray Conference,\\
		30 July- 6 August, 2015\\
		The Hague, The Netherlands}

\begin{document}

\section{Introduction}

Gamma-ray astronomy in the very-high energy range (VHE, E > 100 GeV) can be conducted from the ground using large optical telescopes that detect the Cherenkov light emitted by gamma-ray showers in the atmosphere. Over the last few decades, Imaging Air Cherenkov Telescope (IACT) arrays have been used to study the TeV gamma-ray sky, unveiling more than 150 VHE sources and enabling searches of dark matter and new physics.

The Cherenkov Telescope Array (CTA) is the next generation VHE instrument. It will consist of two telescope arrays, one in each hemisphere, that will provide a full-sky view of the high-energy universe with an order-of-magnitude improvement in sensitivity over the current generation of IACT observatories. Prototype telescopes of various sizes are being designed and constructed for CTA, with optical parameters optimized to increase the sensitivity of the array in different energy ranges. A detailed description of CTA and its capabilities is available in \cite{CTA}.

The energy range between 100 GeV and 10 TeV, at the core of CTA's sensitivity, will be covered by Medium-Sized Telescopes (MSTs) equipped with primary mirrors $\sim$10--12 m in diameter. The CTA-US group is working on the development of a novel type of MST based on the dual-mirror Schwarzchild-Couder optical design. 

The medium-sized Schwarzchild-Couder Telescope (SCT) \cite{SCT} has a 9.6-m primary and a 5.4-m secondary mirror that demagnify the Cherenkov images of air showers allowing for a small plate scale covering a wide field of view of 8$^{\circ}$. The small plate scale enables the use of small-sized and cost-effective silicon photomultiplier (SiPM) detectors, instead of the traditional photomultiplier tubes used in current IACTs. The small size of SiPMs allows for a higher density of photodetectors in the focal plane, providing high-resolution shower images. 
The SCT camera will consist of 177 photodetection modules with 64 SiPM pixels each. These modules are arranged hierarchically into 9 subfields of 5x5 modules each that cover the full field of view of the camera. Given the optical parameters of the SCT, each subfield will cover a field of view of $2.7^{\circ} \times 2.7^{\circ}$. More information about the SCT camera is available in contribution \cite{SCTCam} of these proceedings. 

A prototype SCT (pSCT) equipped with a partial camera of 1600 pixels (i.e. a single subfield) is currently under construction. The assembly and integration of the telescope at the site of the VERITAS IACT array in the F.L. Whipple Observatory in southern Arizona, USA will begin in the Fall of 2015. 

Simulations of the SCT indicate a trigger rate of several kHz due to cosmic-ray showers. The combination of the multi-kHz trigger rate and the large number of pixels in the camera will result in a high data throughput that must be handled by the data acquisition system of the telescope and transmitted elsewhere for further processing or storage. In this work we summarize the current status of the development of the SCT data acquisition system and present a brief outline of future plans.

\section{Data acquisition electronics}

\begin{figure}[tb]
\centering
\includegraphics*[width=0.5\textwidth]{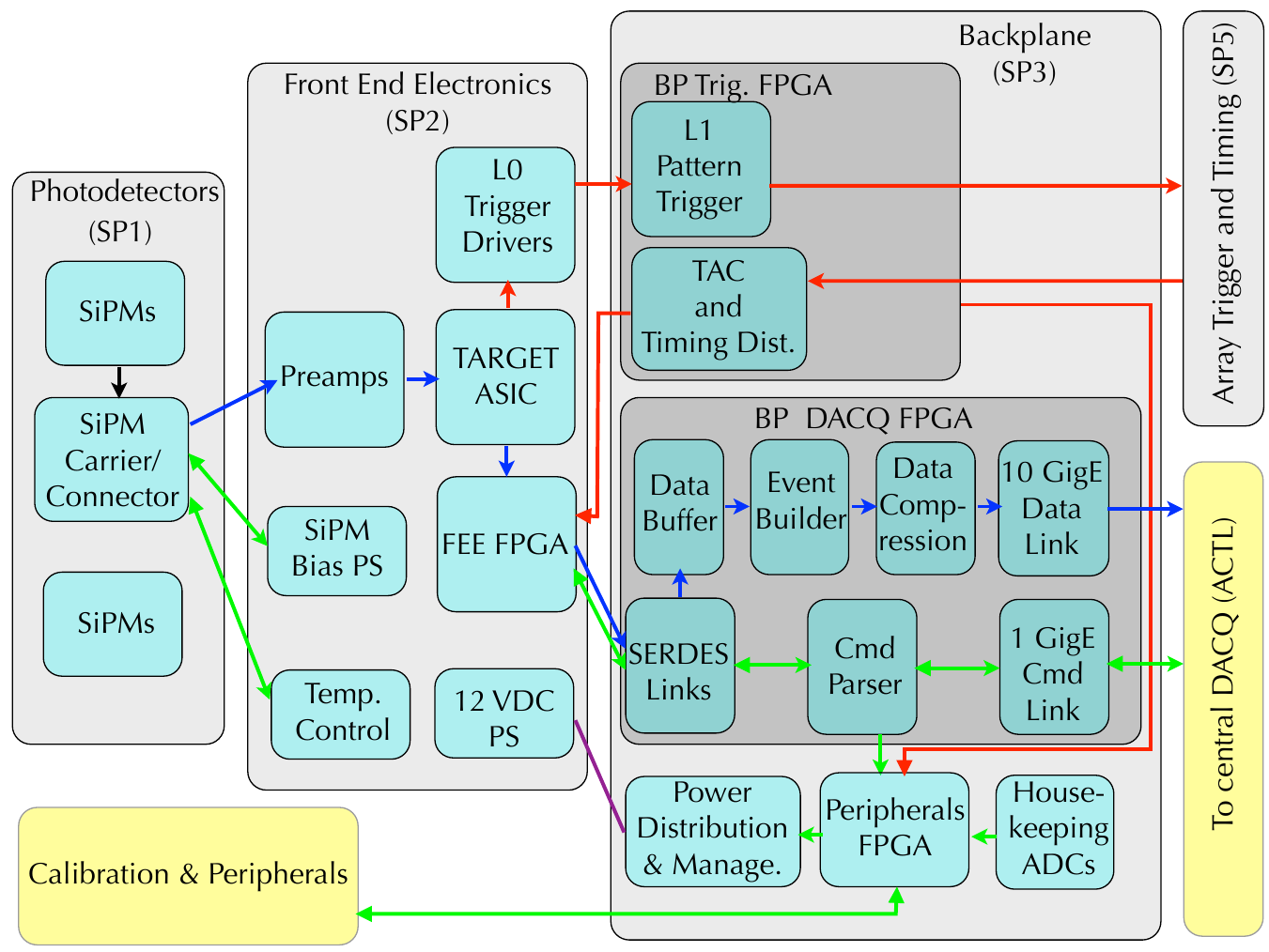}
\caption{Schematic representation of the SCT data acquisition system from the SiPM detectors to the Camera Data Server (indicated here as central DACQ).
\label{daq_sketch}}
\end{figure}

The data acquisition system of the SCT (represented schematically in Fig.~\ref{daq_sketch}) is responsible for transforming air-shower images viewed in the telescope focal plane into data that can be analyzed to reconstruct the direction and energy of the primary gamma-ray photon. The detection of a shower begins with the collection of Cherenkov photons by the telescope which are focused onto the camera, where they are detected by the SiPMs. A detailed description of the SiPMs is available in Refs.~\cite{SiPM1, SiPM2}. 

The signals from the SiPMs are delivered to a set of pulse-shaping preamplifiers, which send the analog signals to the front-end electronic (FEE) modules that contain the custom-designed TARGET Application-Specific Integrated Circuits (ASICs) \cite{TARGET1, TARGET2} that perform the waveform digitization. The TARGET devices have a 300 MHz analog bandwidth and a programmable sampling speed of 0.5 or 1 GSa/s. Efficient triggering is accomplished by summing adjacent image pixels in groups of 4 inside the TARGET ASIC before feeding them to an on-chip discriminator which is the Level-0 (L0) trigger. The 400 trigger signals from the 25 FEE modules in each camera subfield are presented to the Field-Programmable Gate Array (FPGA)  in the camera backplane (BP) that runs the Level-1 (L1) trigger algorithm. The camera BP is described in detail in the following subsection.

\subsection{Camera backplane}

The backplane electronics combine several subsystems into a single monolithic board for each camera subfield.  These functions include data
acquisition, the Level-1 pattern trigger, power distribution and monitoring, housekeeping, and time synchronization and distribution. 

The central data acquisition function of the backplane
is to merge the 25 serial waveform data streams from the FEE modules for an SCT
subfield.  Communications between the FEE electronics and backplanes include
both high-speed waveform data and commands (multiplexed using the PGP\footnote{
The Pretty Good Protocol (PGP) is a VHDL FPGA module developed by the SLAC group for prior experiments that facilitates
bi-directional transmission of framed messages over a two-wire high-speed
SerDes link. \texttt{https://www.slac.stanford.edu/exp/npa/design/pgp\_design.pdf}} protocol), as
well as discrete clock and handshaking signals.  The TARGET-7 ASICs on the FEE
modules buffer the data in a deep analog pipeline, waiting for a trigger signal
(TACK) from the backplane.  The TACK message provides the absolute event time
used to look back in this analog memory and to locate the analog memory cells
corresponding to the time interval of interest, as well as to time-stamp the event with a unique number (for event building).  On receipt of a TACK message,
a state machine in the FEE FPGA initiates digitization, formats and writes the
waveform data into the internal FEE FPGA FIFOs.  The FEE modules act as \emph{servers}
waiting for TACK messages (array trigger signals), and act as \emph{clients} with
respect to the backplanes, pushing data whenever the FEE write FIFOs have data.

Synchronization of the FEE and
backplane clocks is accomplished by the 
DIAT system (see \cite{DIAT} for details) through an interface connector and
precision clock distribution components on the
backplane.  The backplane 125-MHz clocks are phase-locked to the common 62.5-MHz 
oscillator signal from the DIAT and distributed to the FEE modules.  Local
counters on the backplane interpolate the 125-MHz clock to 1 ns.  On the FEE
modules, the 125-MHz (8-ns) clock is the fundamental digital time reference,
but a delay-locked 1-ns domino clock inside the TARGET ASICs is kept in phase 
with the 8-ns clock to $\sim$100~ps (much less than one sample time).
The FEEs receive the 64-bit 1-ns clock in the TACK message for inclusion in the data records. 

In the end, all counters and clocks on the different FEE modules and backplanes are
synchronized (with occasional imperative messages to correct synchronization)
so that the FEE data can be localized to 1-ns resolution with respect to the
common array trigger.  

The 1-ns time (encoded in the TACK message) 
is used as the unique identifier (inserted
into the data records by the FEE) for event building on the backplane.
An event counter maintained in the FEE FPGAs is included in the
data as a redundant mechanism for event building and as a check for lost data.

The backplane DAQ boards include FPGAs that run a server ready to receive the data through
the PGP SerDes links.  Inside the backplane DAQ FPGAs, these asynchronous data
streams (wrapped in UDP packets) are unwrapped, parsed and assembled into event
data by the Level-1 event builder (a Verilog block inside the DAQ FPGA).  Data
are also buffered in the DAQ board to average out the data rate and allow
block transfer to the telescope data acquisition system.
Several alternatives are being considered for the physical realization of these
DAQ boards.  While initially we may make use of existing open-source FPGA
protoboards, we plan to make use of a board
designed by Seven Solutions\footnote{http://www.sevensols.com/} for the CHEC camera group in CTA.
These boards were designed in
consultation with the Washington University group to provide signal and connector compatibility
with the SCT backplane.  

Currently the boards include two 1-Gb/s Ethernet transceivers for the data link, while
the nominal design calls for a 10-Gb/s link. This modification can be implemented by the manufacturer if necessary. 

\begin{figure}[tb]
\centering
\raisebox{-0.5\height}{\includegraphics*[width=0.52\textwidth]{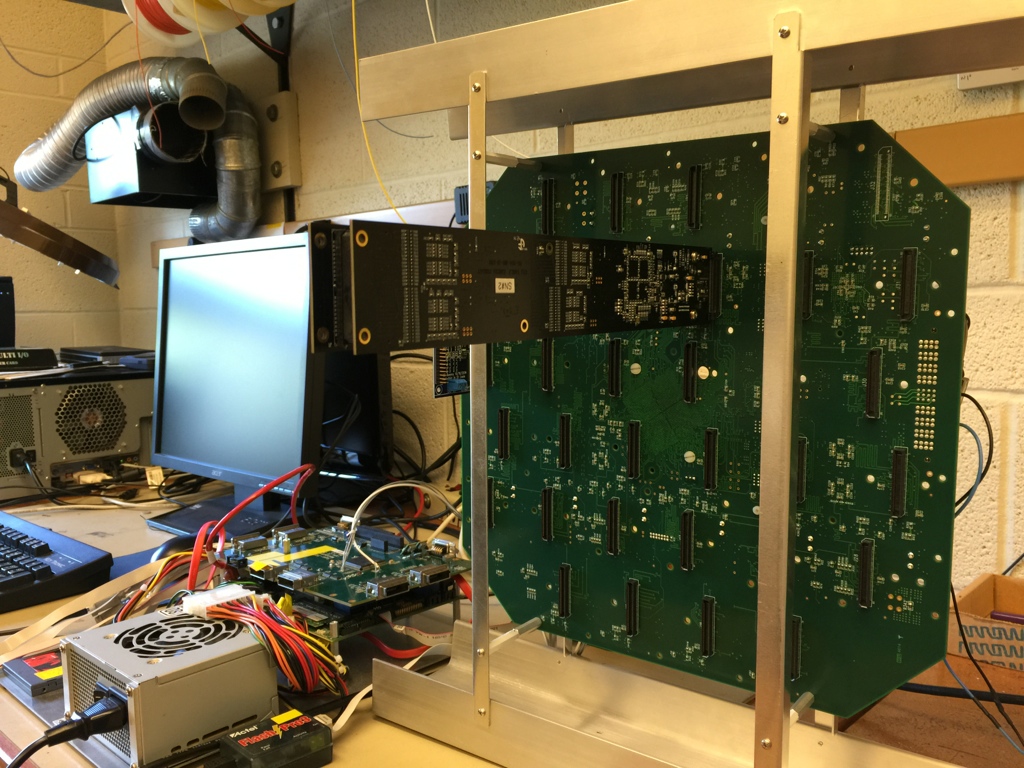}}
\caption{Backplane with Target 7 FEE module in a test setup at Washington University in St. Louis.
\label{bp_fee_test}}
\end{figure}

The Level-2 (L2) trigger collects L1 triggers from the backplane and provides other timing signals required for event synchronization, event tagging, clock synchronization, and synchronous resets.  The L2 will have the trigger records from all backplanes and can issue trigger commands based on the entire camera and force readout of all backplanes. If required, the events built in the backplane can be zero-supressed and compressed in the backplane FPGA before transmitting them to alleviate the load on the data links to the Camera Data Server (CDS). The Camera Data Server is described in the next subsection.

Another major function of the backplane is to provide the command and slow
control interface to a number of the camera's electronics subsystems.  UDP
packets received by the DAQ FPGAs are decoded to determine if they are
commands, then directed either to the housekeeping FPGA (an ACTEL
FPGA on the backplane dedicated to slow control and monitoring), FEE modules,
or trigger FPGA.  Commands can be issued to set data acquisition parameters in
the FEE modules, set trigger parameters, issue slow control commands and to
read out various bits of status information.   The housekeeping FPGA controls power-up
sequencing, emergency power shutdown to FEE modules, monitors currents and
voltages using ADCs on the backplane, and provides an interface to a general
purpose peripherals connector.

The final function of the BP is to distribute power to the FEE and SiPM
modules.  Power for the FEE PCBs and Peltier coolers is provided by an external
12 VDC power supply and distributed by the backplane through high-current bus
bars and PCB traces or segmented planes to the FEE/BP connectors.  High voltage
power for the SiPMs is provided through an external (programmable) power
supply, and distributed through separate PCB traces to the FEE/BP connectors.

One backplane is being used for FEE integration and mechanical tests in Washington University (see Fig.~\ref{bp_fee_test}). 
The FPGA code is being developed and integration with the prototype Camera Data Server will take place before 
Fall 2015. 

\subsection{Camera Data Server}

The Camera Data Server (CDS) is the computer that will receive the event data stream from all 9 backplanes in the SCT (one in the pSCT) and will provide high-level event building and time-stamping. The events built in the CDS will be formatted according to the specifications defined by CTA before sending them downstream to other processing nodes for analysis or storage. 

A key quantity for the selection of the CDS components is the data rate that will be received from the telescope. The data acquisition electronics are designed to handle a single telescope trigger rate up to 10 kHz. However, a suppression factor of at least 4 is provided by the DIAT system, leading to an effective single-telescope trigger rate of 2.5 kHz.  For a full-camera readout size of 1.42 MB (125 bytes/pixel for 11,328 pixels) this trigger rate translates into a raw data rate of 3.54 GB/s (28.3 Gb/s) per telescope.

The data connection between the SCT camera and the CDS will consist of a series of optical fiber lines. Two fiber lines from the Seven Solutions DACQ board in each of the 9 backplanes (i.e. a total of 18 fibers) will be connected to a high-speed switch at the base of the telescope. From there, data will be transmitted directly to the CDS located in the CTA data center using the four 10-Gb/s fiber lines allocated for each SCT in the CTA infrastructure planning. The data produced by each telescope (28.3 Gb/s) can be transmitted over these four lines but, if additional reduction is needed, options may be available to perform data compression, zero suppression, or waveform integration in the camera FPGAs prior to transmission. 

The camera data will be transmitted to the CDS as UDP packets that will contain the event waveforms. Each waveform has 60 samples with 2 bytes per sample and a 5-byte header. The size of the UDP packet will be determined in throughput studies of the CDS software, currently under development. Upon arrival at the CDS, each UDP packet is parsed and a partial event is assembled based on the information contained in the packet. These event fragments are indexed according to their global event number and put in a buffer to wait for the arrival of all event fragments from different backplanes. The ordering of the fragments in the buffer follows the time of their arrival at the CDS. The depth of this buffer is approximately 0.5 seconds, which at the expected SCT trigger rate results in a buffer size of 1.8 GB. 

A separate thread will be in charge of building complete events from the fragments stored in the memory buffer. After a certain time, the first (i.e. the oldest) fragment in the buffer is selected and its global event number is used to scan the buffer for all fragments that correspond to the same event. The complete event is put in a second buffer to await transmission or storage and the event fragments are removed from the first buffer. The format for transmitting camera events to the central CTA data acquisition system to form array-level events is currently being defined. One of the possible ways to transmit this data is by making use of the Protocol Buffers serialization library from Google \footnote{
https://developers.google.com/protocol-buffers/}. For the pSCT, we expect to use the VERITAS format for data storage to allow merging the data from the telescope with that taken at the same time with the VERITAS array. This will enable the study of stereo events and the validation of the pSCT performance.

For the prototype CDS, we plan to use the Dell PowerEdge R730xd rack server 
which is being considered for other camera servers as a cost-effective solution that provides good flexibility for different server configurations. 
 

The data acquisition system will be integrated into the CTA array control system with well-defined interfaces to the common data and control flow.  For the local command and control system, we assume a virtual bus architecture implemented with a separate 1-Gb/s network, based on a publish/subscribe design philosophy.  Individual camera systems will each have a server with an Ethernet interface that allows the receipt of commands to change settings, and can publish house-keeping telemetry.  Client computers can subscribe to the telemetry (and can provide graphical monitoring for individual systems, an entire camera, or the CTA array) and can (through the use of a lock mechanism) take over control of the command link to the subsystems.  The command computer that issues commands (and subscribes to telemetry) may be a local computer system or central system that is part of the observatory.  

\section{Future plans}

After the ongoing backplane tests in Washington University, work will concentrate on software development for the camera server and its integration with the backplane. Throughput tests are planned for the camera server software to optimize the size of the UDP packets and buffers and to test the performance of the VERITAS file format for pSCT data storage. The software and hardware components of the DAQ system will be available for the tests and final installation of the prototype SCT at the VERITAS site, which will take place in Fall 2015.

\section{Acknowledgements}

We gratefully acknowledge support from the agencies and organizations listed under Funding Agencies at this website: http://www.cta-observatory.org/. The development of the prototype SCT has been made possible by funding provided through the NSF-MRI program.

\end{document}